# AFM-based Hamaker Constant Determination with Blind Tip Reconstruction

*Benny Ku[1,2], Ferdinandus van de Wetering[2], Jens Bolten[3], Bart Stel[2],*

*Mark A. van de Kerkhof[2], Max C. Lemme[1,3]*

[1] Chair of Electronic Devices, RWTH Aachen University,

Otto-Blumenthal-Str. 25, 52074 Aachen, Germany

[2] ASML Netherlands B.V., De Run 6501, 5504 DR Veldhoven, The Netherlands

[3] AMO GmbH, Advanced Microelectronic Center Aachen (AMICA),

Otto-Blumenthal-Str. 25, 52074 Aachen, Germany

Keywords: adhesion force; atomic force microscope (AFM); blind tip reconstruction; Hamaker theory; EUV photomask particle contamination

Particle contamination of extreme ultraviolet (EUV) photomasks is one of the numerous challenges in nanoscale semiconductor fabrication, since it can lead to systematic device failures when disturbed patterns are projected repeatedly onto wafers during EUV exposure. Understanding adhesion of particle contamination is key in devising a strategy for cleaning of photomasks. In this work, particle contamination is treated as a particle-plane problem in which surface roughness and the interacting materials have major influences. For this purpose, we perform vacuum atomic force microscopy (AFM) contact measurements to quantify the van der Waals (vdW) forces between tip and sample. We introduce this as a vacuum AFM-based methodology that combines numerical Hamaker theory and Blind Tip Reconstruction (BTR). We have determined the Hamaker constants of $15 \times 10^{-20}$ J and $13 \times 10^{-20}$ J for the material systems of a silicon (Si) tip with both aluminum oxide ($Al_2O_3$) and native silicon dioxide ($SiO_2$) on Si substrates, respectively. Our methodology allows an alternative, quick and low-cost approach to characterize the Hamaker constant within the right order of magnitude for any material combination.

# 1. Introduction

The introduction of extreme ultraviolet (EUV) lithography for high volume manufacturing in the semiconductor industry was a major breakthrough in the recent years.[1] Nevertheless, like any technical system, EUV scanner systems continue to benefit from performance optimization. Chip fabrication with critical dimensions below 10 nm imposes increasingly stringent requirements regarding contamination in comparison to former photolithography generations. EUV photomasks must often be replaced once they become critically contaminated. This is a time-consuming process that leads to downtimes of the EUV scanners.[2] Furthermore, EUV photomask fabrication is very complex and financially cost-intensive.[3] Thus, there is substantial interest in cleaning contaminated EUV photomasks.[4] Several different EUV mask cleaning methods exist, depending on the type of contamination[5], among which plasma lifting in vacuum conditions is a very promising one.[6]

The main source of EUV mask contaminants are solid particles.[2] The interaction between such particles and the mask surface is quantified by their adhesion force, which results from the sum of different attraction forces, namely van der Waals (vdW), capillary, electrostatic and chemical forces.[7] A successful cleaning method requires a release force that can overcome the adhesion force between the contamination particle and the photomask surface.[8] It can be quantified experimentally by releasing particles with known forces. A reliable prediction of those release forces with a suitable model offers efficient guidance for such experiments, minimizing the overall experimental effort.

To quantify the amount of vdW forces within the adhesion force, such modelling can be based on the Hamaker theory.[9] It approximates the interaction between two solid bodies. Within this theory, the Hamaker constant ($A_H$) is a key parameter that incorporates the material properties into the interaction, such as optical constants, dielectric constants and electron density oscillations (i.e. the plasma frequency or Langmuir waves).[10] Substantial work has been carried out to define the Hamaker constant for commonly used materials, and the results can be found in databases today.[10,11] The Hamaker theory does not consider solid-state properties of many-body systems, where electron screening plays a role. This issue has been solved by introducing the quantum field theory, or Lifschitz theory.[12–14] However, in the specific application of EUV masks covered here, the contamination particles are typically sufficiently large (with greater than 50 nm diameter)[2] to ignore quantum effects in the description of their adhesion. Therefore, the Lifschitz theory does not need to be applied for this specific class of problems.

In addition to the Hamaker and Lifschitz theories, several other models address the vdW interaction for a sphere-plane problem. There are models based on contact mechanics[15–17], analytical vdW models with averaged roughness statistics[18,19] and numerical models based on atom-atom interaction.[20,21] However, these models have limitations either in their applicability or their computation time for the specific application of EUV photomask contamination. For instance, analytical models often assume a spherical-shaped particle with additional roughness, but it is not generally accurate to approximate any given particle as a sphere. Moreover, EUV photomask contaminant particles are too large to use an atomistic model, as this would require excessive computation time.

Experimental methods to extract the Hamaker constant through atomic force microscopy (AFM) are based either on the dynamic mechanical properties of the AFM cantilever[22–24], idealized (smooth surface) assumptions[24–26], or on scanning electron microscopy (SEM) or transmission electron microscopy (TEM) data of the AFM tip.[27–29] Idealized models can be used where they can sufficient approximate a real system, but they become limited in their accuracy once surface roughness is involved, which is the case for most real systems.[30] Other methods have a risk of tip deformation or oxidation during TEM preparation, so that the TEM data does no longer accurately represent the tip morphology at the time of the actual AFM measurements.

Although research on adhesion forces have been conducted for many decades already, they are not fully understood yet, as recent studies show.[31–33] Characterizing adhesion forces between two arbitrary-shaped bodies as detailed as possible is still an active field of research. In this work, we present an alternative quick and low-cost methodology, which takes into account surface roughness and body shape on the nanoscale. Thus, we use the geometry of the AFM tip and surface roughness instead of a statistical parameter for our model. Besides particle contamination removal, knowing the adhesion force is beneficial for applications fields such as biotechnology[34], micro-electro-mechanical systems (MEMS)[35] or 2D materials.[36]

In particular, we develop a methodology which combines the numerical Hamaker theory in the form of pairwise summation[27,37] and blind tip reconstruction (BTR) to compute the vdW force.[38–41] This methodology is verified with experimental data and offers in-situ insights immediately before and after the force measurement within a vacuum AFM. The surface roughness of a sample is obtained through AFM by scanning the sample topography. Then, the tip morphology data from the AFM image is obtained by BTR, which uses an algorithm to mathematically deconvolute the sample surface dataset and reconstruct a 3D model of the tip shape.[40] This is possible because the scanned topography image can be considered as the fingerprint of the AFM tip. The experimental data of particle and surface is compared with the numerical vdW force model, i.e. by comparing force histograms and determining a reasonable Hamaker constant. Although this methodology does not aim the best accuracy, it still offers a fast, simple, and low-cost workflow to determine the Hamaker constant of any material system within the right order of magnitude. Usually, knowing the right order of magnitude of the Hamaker constant of interest is sufficient for particle removal applications, as applied removal forces only need to overcome the adhesion force. However, the removal force should not be too large either, otherwise this could damage the mask surface

The vdW force model detailed here requires information on the morphology of the particle and substrate surface. In the literature, randomized rough surfaces have been generated by various roughness models.[42,43] In our work, the vdW force model has been combined with fractal surfaces[44], which adds flexibility in generating and characterizing surface topographies.

In fractal theory, the surface roughness is defined by two key parameters: One is the roughness height root-mean square (rms) $h_{rms}$, which is an indication factor for the amplitude of the roughness asperities. The other one is the Hurst exponent $H$, which represents the exponential decay rate of the roughness asperities, thus indicating whether a surface is smooth or rough.[44] Based on those two key parameters, the topography length $L$ and the number of datapoints $N$ of fractal surfaces can be characterized through their power spectral density (PSD) by a parameter set of ($h_{rms}$, $H$, $L$, $N$). This property is used to generate surfaces,[45] which can be further refined with random phase shifts[46] to randomize the height distribution without changing the PSD. This definition of the rough surface is typically featured in the surface function of the plane ($Z_B$) but can also be used to create rough particles in the surface function of the particle ($Z_A$), which are both defined in the Hamaker theory. Additional detailed information about the applied vdW force model is summarized in the Supporting Information.

## 2. Methodology

The general workflow for obtaining $A_H$ using the proposed methodology and model is shown in Figure 1. Information about the sample topography and surface forces on $N_{exp}$ data points are all obtained by contact mode, as it provides the best measurement signals. Operation in vacuum reproduces the typical conditions used for plasma lifting, a promising method for EUV mask cleaning. In addition, the vacuum prevents immediate tip oxidation if the tip is based on Si and should get damaged during the contact measurements. The method requires two samples to be placed on the movable stage inside of the AFM vacuum chamber: A *MikroMasch PA-01* reference sample used to characterize the tip by means of BTR, and a measurement sample used for the actual extraction of the Hamaker constant for the specific tip/sample material combination. In this work, a commercial vacuum AFM system *NX-Hivac* by *ParkSystems* has been used at a vacuum pressure of $10^{-4}$ mbar. The tip model used for the experiments was *PPP-NCSTR*, which is based on Si.

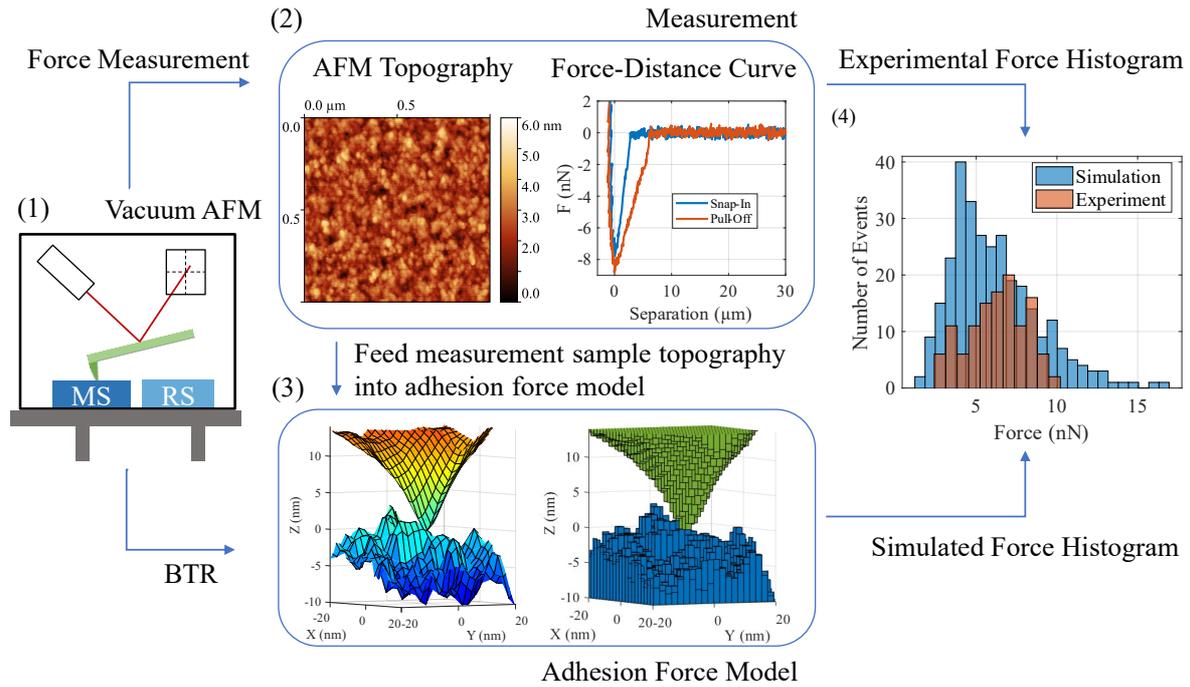

**Figure 1.** Workflow overview of the AFM measurement, the experimental data, the vdW force model and the force histogram evaluation. (1) A vacuum AFM is used to scan the surface topography of the measurement (MS) and the reference sample (RS). (2) The surface topography of both samples can be analyzed, as well as the pull-off forces from the series of force measurement. (3) The surface topography scans of the reference sample are used to reconstruct the tip profile by BTR, which can then be fed into the vdW force model. (4) The simulated histogram is fitted to the experimental one with a suitable Hamaker constant.

The experimental part of the methodology can be divided into the following steps:

(1) Performing a topography scan on the reference sample for a BTR characterization of a new AFM tip. Topography scan settings have been fixed at (1×1) μm$^2$ scan area and (512×512) data points for all measurements.

(2) After AFM cantilever calibration, which is required to find the right force constant/sensitivity for the cantilever, the BTR characterization is repeated by scanning the reference sample again.

(3) The force measurement series are performed. Each measurement has been conducted in 100 nm steps on a (1×1) µm² scan area, thus resulting in a total of 121 measurement points per measurement. The experimental data of the pull-off forces of the series of force-distance curves can be collected in a histogram, representing the statistical behavior of the vdW force between tip and substrate material. Despite the measurements in vacuum, pull-off forces still show residual capillary forces, which were sometimes observed during the measurements (see Supporting Information).

(4) The topography of the reference sample is scanned again for a BTR characterization after the force measurements.

(5) The topography characterization of the measurement sample is performed independently from the force measurement series with a second and new tip. This minimizes measurement errors in the topography data. Then, roughness data, namely $h_{rms}$ and $H$, can be extracted from the measured topography of the force measurement sample. These data can then be used to generate a series of simulated force histogram with $A_H$ as a fit parameter.

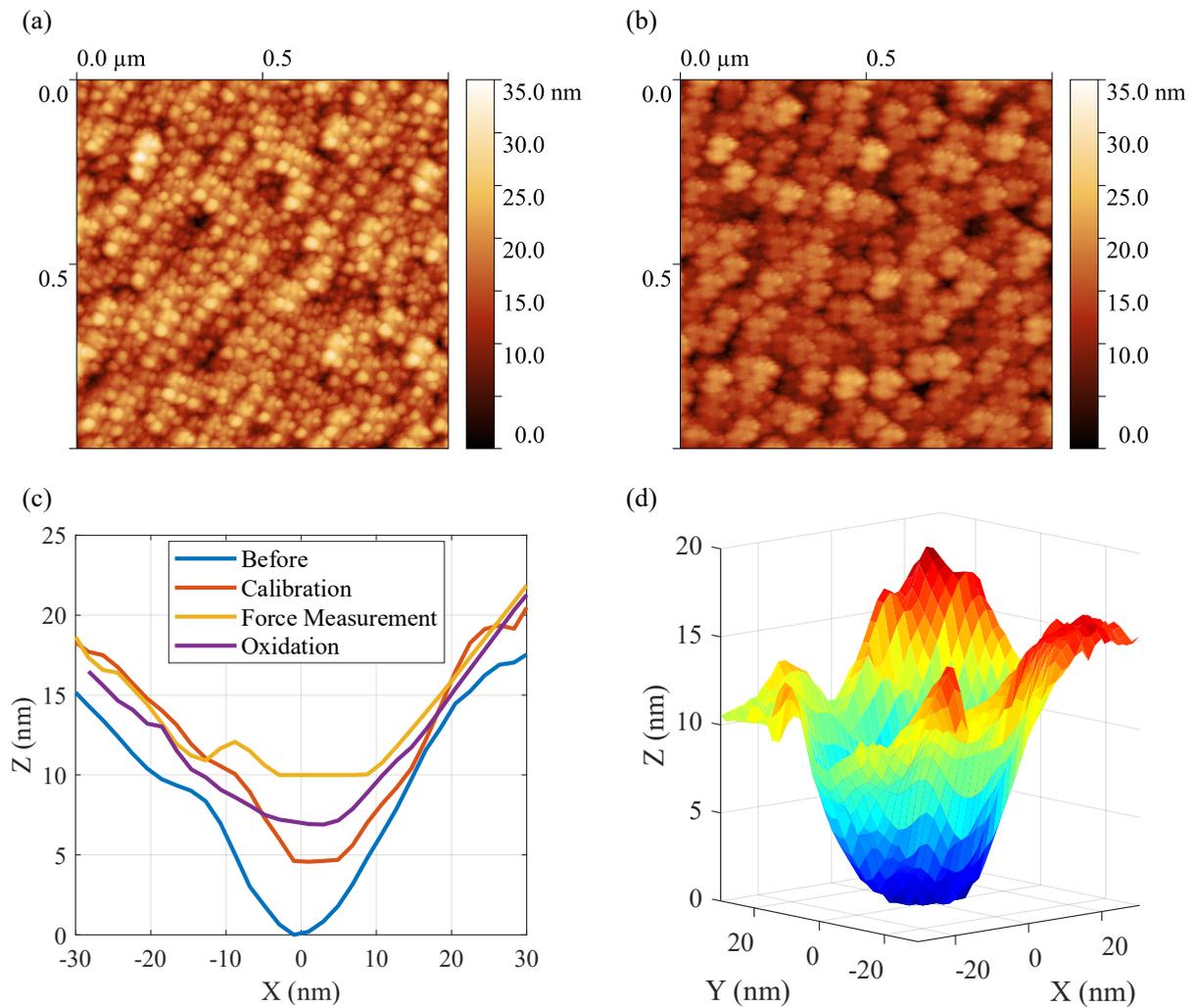

**Figure 2.** Topography scan of the reference sample *MikroMasch PA-01* (a) before and (b) after the force measurement. Loss in sharpness and detail of the scan image indicate an alteration of the AFM tip. (c) Height profiles of a reconstructed AFM tip along the *X* axis in chronological order: at the beginning (blue), after tip calibration (red), after force measurement (yellow) and after oxidation (purple). (d) 3D profile as reconstructed by BTR of the tip shape after a force measurement (yellow curve in Figure 2c).

The vdW force model requires knowing both the surface topographies of the BTR and the force measurement sample as well as their geometrical remodeling. To accurately characterize the tip using BTR and provide the data needed for modelling, the topography of the reference sample needs to fulfill some requirements. First, it requires sufficiently large topographical structures to process the tip characterization by BTR. These surface structures can have any form, e.g. arbitrarily formed nanoparticles or pyramidal structures, which are distributed over the whole reference sample surface. In the case of vdW interactions, the structures should have peak-to-valley distances of at least 20 nm to obtain tip morphology data within the whole vdW interaction range. Here, we used a *MikroMasch PA-01* reference sample with pyramidal structures with heights of up to 50 nm for the characterization by BTR. Data of topography scans performed in vacuum before and after contact measurements are shown in Figure 2a and 2b, respectively. The difference in the topography scans can be attributed to changes of the tip geometry between measurements, as illustrated in the tip-shape profiles extracted by BTR in Figure 2c. The plots show the height profile along an arbitrary $X$ axis of a tip in different conditions., initially (blue), after tip calibration (red), after force measurement (yellow) and after oxidation (purple). Note that the Z-axis in Figures 2c is arbitrarily shifted to enhance the clarity of the graphs. In Figure 2d, the tip profile reconstructed by BTR after a force measurement (topography scan in Figure 2b yellow curve in Figure 2c) is shown as a 3D plot. There are small trenches between the pyramidal structure on the reference sample, from which the BTR algorithm obtains the smallest possible tip profile for the reconstructed tip shape.

It should be noted that the topography image of a deformed tip has image noise, so the BTR algorithm may need a too larger threshold value to exclude the image noise. This can lead into a flat section of the reconstructed tip, where some atomistic details are lost (Figure 2c). Nevertheless, a sufficient approximation of the tip geometry can still be reconstructed. Here, a tip diameter of roughly 40 nm was obtained. The simulated geometrical data of the 3D profile can then be used as input data for the vdW force model.

In our approach, the total vdW force is obtained by numerically integrating the vdW forces across the surfaces of the tip and the substrate. This is achieved by dividing both the geometrical surface data of the AFM scans and the BTR data into small elements and integrating over these segments. This is illustrated in Figure 3 with the example of an idealized sphere particle and a rough plane surface, that have been divided into rectangular pillars. Here, rectangular pillars have been chosen, as they can be efficiently computed in a Cartesian coordinate system. Moreover, the force interaction between two rectangular pillars has been analytically formulated by the Hamaker theory.[9] In the example, a spherical particle and an (1×1) µm$^2$ artificially generated surface topography have been reduced to the area contributing to the vdW forces, i.e. only part of the sphere was used (Figure 3a). Based on these data, the areas were clustered into rectangular pillars that represent a meaningful density of datapoints *Z(X,Y)* across both areas of the sphere and the surface (Figure 3b). The vdW force of each pillar of the sphere can then be calculated with respect to each pillar of the surface.

The roughness of a sample typically shows statistical variations. The experimental and simulated data thus has a significant variance with a certain spread around the most probable force value. Comparing them, as anticipated in the workflow in Figure 1, therefore requires statistically relevant data sets. These were obtained by scanning the AFM tip across a reasonably large number of random measurement spots and collecting the data in a force histogram, which then accurately represents the force statistics, including minimum and maximum values and the mean adhesion force. The relation between the surface roughness and the force distribution was investigated by simulating the vdW force interaction between an idealized sphere and an artificial surface with random roughness (Supplementary Information Figure S1).

(a) (b)

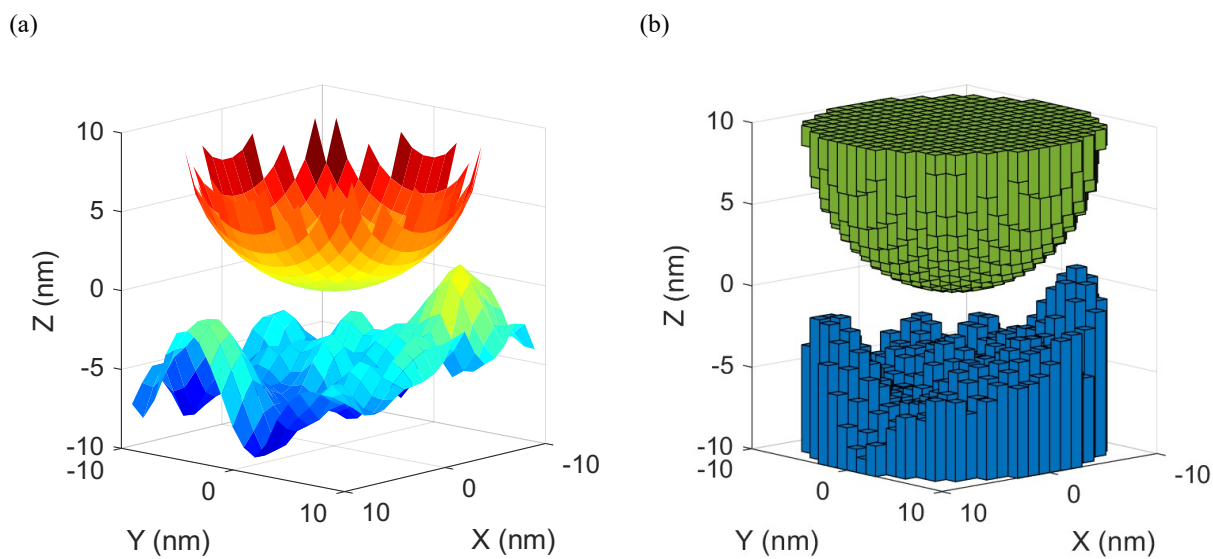

**Figure** 3. (a) Local simulation domain of an ideal spherical particle on a surface with randomly generated surface roughness. (b) Clustered representation in the vdW force model. The surface underneath the particle is reduced to its projected area.

## 3. Results and Discussion

The proposed concept of the AFM-based methodology that combines BTR with the vdW force model has been validated with two substrates, $Al_2O_3$ and $SiO_2$ on Si. AFM measurements showed a roughness of $h_{rms} = 0.06$ nm and $H = 0.14$ for the $Al_2O_3$ substrate and a roughness of $h_{rms} = 0.77$ nm and $H = 0.5$ for the native $SiO_2$ on Si substrate (Figure 4a and 4b). The AFM topography scans of both samples are plotted in the same color scale to highlight their significant difference in roughness. As the AFM tips used in this study are based on Si, a native oxide layer may form on the tip and affect the force calculations. We have therefore considered both tip materials, Si and $SiO_2$.

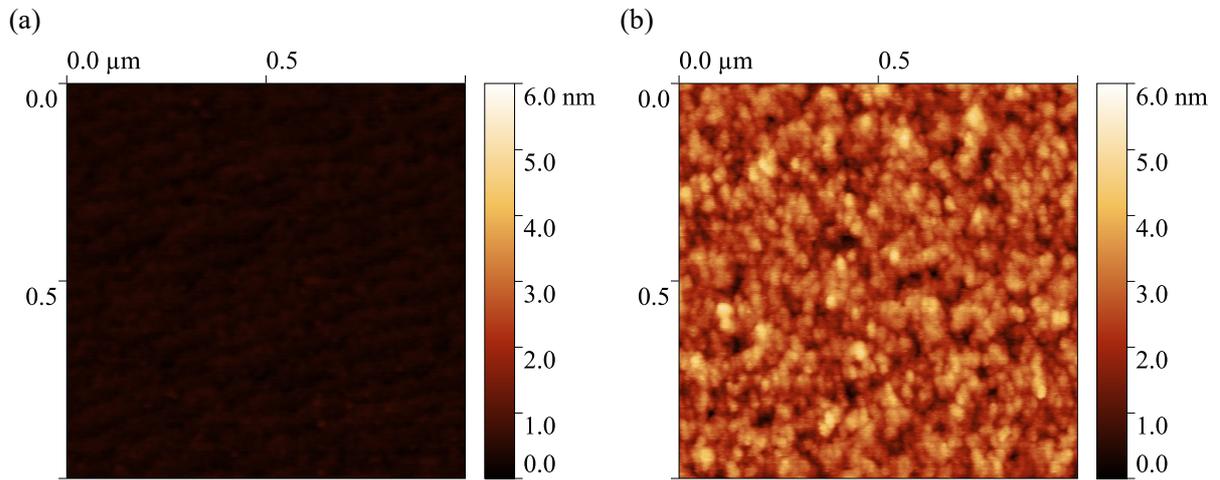

**Figure 4.** Topography scan of force measurement samples. The Z-scale has been aligned from 0 nm to 6 nm to highlight the difference in roughness between both samples. (a) $Al_2O_3$ substrate with $h_{rms} = 0.06$ nm and $H = 0.14$. (b) $SiO_2$ on Si substrate with $h_{rms} = 0.77$ nm and $H = 0.5$.

**Table 1.** Hamaker constants taken from literature[7] define the expectation range for the material systems in the experiment. The last column shows the Hamaker constant extracted by the methodology proposed in this work.

| Sample (1) | Tip (2) | $A_H^{(11)}$ | $A_H^{(22)}$ | $A_H^{(12)}$ | Experimental $A_H^{(12)}$ |
|---|---|---|---|---|---|
| Al$_2$O$_3$ | Si | 15×10$^{-20}$ J | 20×10$^{-20}$ J | 17×10$^{-20}$ J | 15×10$^{-20}$ J |
| SiO$_2$ on Si Substrate | Si | 7×10$^{-20}$ J | 20×10$^{-20}$ J | 12×10$^{-20}$ J | 13×10$^{-20}$ J |

Sometimes, a small capillary effect still occurred during the measurements in vacuum due to residual nanolayers of water (Figure S2), which resulted in an additional contribution to the pull-off force. This leads to an overestimation of the Hamaker constant, but also a lowering of Hamaker constant due to native SiO$_2$ on the Si tip and substrate (Figure S3). The Hamaker constant for SiO2 is $A_H$ = 7×10$^{-20}$ J, but it was not possible to quantify the exact amount of native oxide on the Si tip.

The experimental force histograms were obtained by extracting the pull-off forces from a series of force-distance curves. Then, force histograms were simulated with the vdW force model, which is based on the BTR characterization data of the tip and the topography of the measurement sample. We used literature Hamaker constants for the substrates of $A_H$ = 15×10$^{-20}$ J for Al$_2$O$_3$, and $A_H$ = 7×10$^{-20}$ J for SiO$_2$, respectively. For the Si-based AFM tips, we used a Hamaker constant of $A_H$ = 20×10$^{-20}$ J. The numeration (1) and (2) indicates the sample and tip material. (11) stands for mixing the Hamaker constants in an interaction of a two-body pair of the same substrate material, while (22) is the respective case for the tip material. (12) is then the cross-action between both sample and tip materials, for which the mixing rule for the Hamaker constant

$$A_H^{(12)} \approx \sqrt{A_H^{(11)} A_H^{(22)}}$$

has been applied to calculate the total Hamaker constant.[7] We have summarized all possible material combinations in Table 1.

By varying $A_H$ in the vdW force model, the simulated force histogram can be fitted in integer steps to the experimental data obtained from the force measurements, until both histogram mean values have the best agreement. For force calculation, it has been decided to use contact distance of $d_0 = 0.3$ nm as tip-sample distance (see the Supporting Information).[47] An example of a simulated force histogram for an $Al_2O_3$ substrate, fitted to the respective experimental force histogram, is shown in Figure 5a. The Hamaker constant of the $Al_2O_3$ substrate determined from this experiment is approximately $15 \times 10^{-20}$ J, which is a sufficient good agreement with the expected value for the literature Hamaker constant of $17 \times 10^{-20}$ J. Figure 5b illustrates that the variability in the simulation data for the measurement on the $Al_2O_3$ substrate is significantly larger than in the experiment. It also shows that the experimentally obtained forces are within a robust statistical margin. This indicates that the tip deformation caused by each scan of the substrate surface in contact mode can be rather neglected. Moreover, Figure 5b shows that the AFM tip is not encountering serious contact charging effects. Hence, the initial assumption of the methodology that such effects can be neglected is confirmed. The relative standard deviation provides an indication for the histogram spread. It amounts to 5.1% for the experimental and to 12.9% for the simulated force histogram.

For all statistical evaluations, we have applied Gaussian statistics to calculate the mean value and standard deviation of the force distributions.

The corresponding force histograms for the $SiO_2$ on Si substrate is shown in Figure 5c. The simulated data was fitted to the experimental data with a Hamaker constant of $A_H = 13 \times 10^{-20}$ J, which is in good agreement of the expected $12 \times 10^{-20}$ J based on literature data. The simulated and experimental histograms show a larger relative standard deviation of 19.5 % and of 41.3 %, respectively.

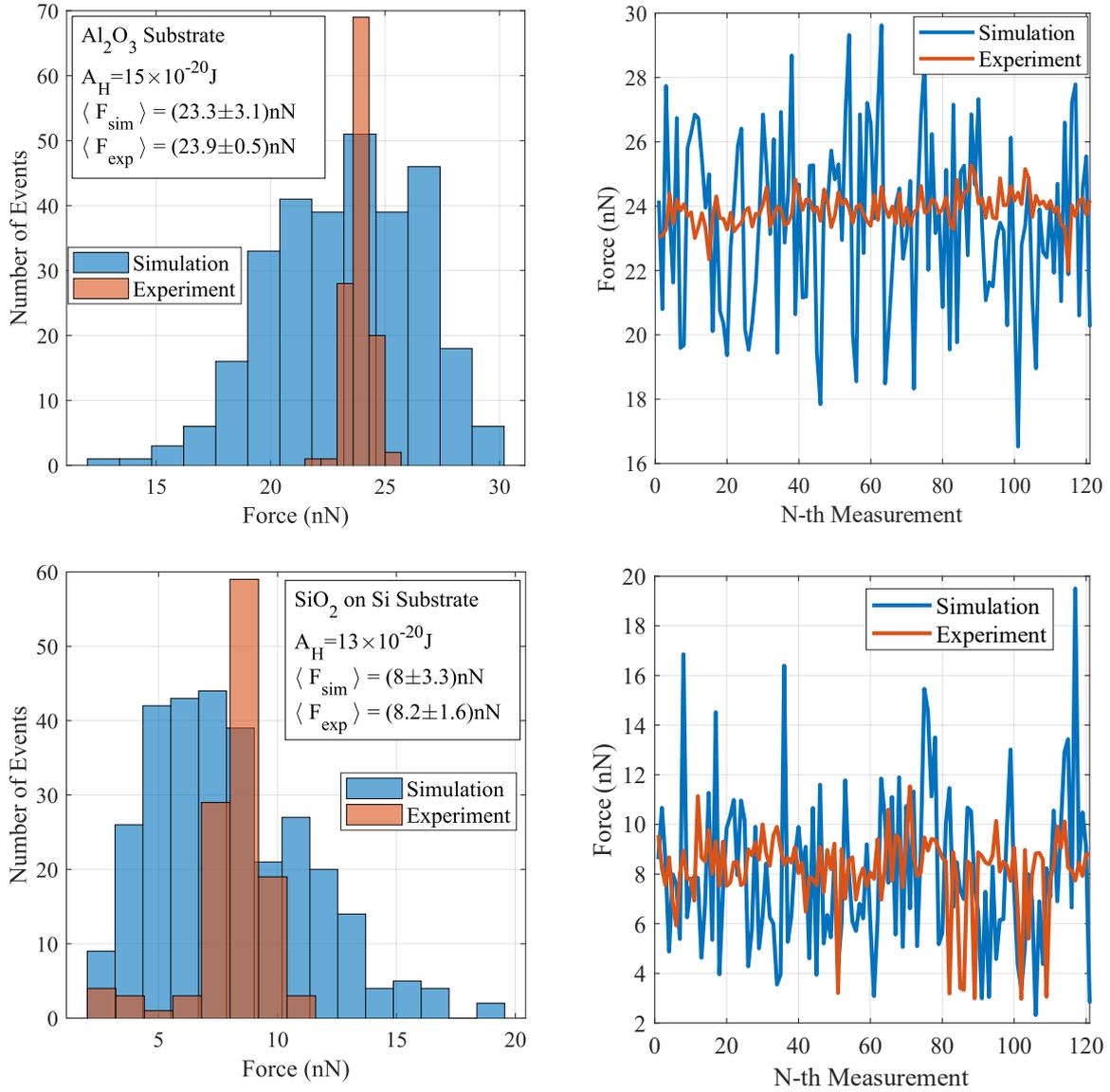

**Figure 5.** (a) Simulated (blue) and experimental (red) force histograms and (b) Simulated (blue) and experimental (red) force at the *N*-th measurement for each force measurement on Al$_2$O$_3$. (c) Simulated (blue) and experimental (red) force histograms and (d) Simulated (blue) and experimental (red) force at the *N*-th measurement for each force measurement on SiO$_2$ on Si. To obtain a robust statistic, 121 force measurements have been conducted for each force histogram. The simulation settings are fixed at pixel size $\Delta L$ = 1.95 nm, tip-sample distance $d_0$ = 0.3 nm and force calculations on 300 randomly selected spots. The experimental force histogram lies within the bandwidth of the force histogram simulated by the vdW force model.

The significant difference between the statistical spread of simulation and experimental data is somewhat counterintuitive. Usually, simulation data shows an ideal and precise result, while the experimental data contains an uncertainty due to effects not accounted for in the modelling and/or measurement errors. Here, the smaller statistical spread in the experimental force histogram can be attributed both to surface deformation and to different surface roughness values. Firstly, the AFM tip is expected to partially deform (indent) the sample surface in alignment with the tip shape as it reaches the sample surface.[15,48] Hence, the tip-sample interaction in the experiment is effectively smoother than in the simulation where this effect is not implemented. Secondly, in the case of the $SiO_2$ on Si sample, the large spread in the experimental force histograms can be explained as a statistical effect of the surface roughness. A rough surface can be defined through crest and trough regions. The adhesion force then strongly depends on the random position of the tip-sample contact, i.e. there is a large difference between a tip-crest adhesion force and that of a tip-trough interaction. This leads to a higher statistical variance for a higher $h_{rms}$, e.g. as shown in Figure 4b.

## 4. Conclusions

Developing suitable methods for particle removal from EUV photomasks requires a fundamental understanding of particle adhesion forces. Here, we proposed an AFM-based methodology with a numerical vdW force model and BTR to determine the Hamaker constant of any given material combination. While the approach is general, its validity has been demonstrated with two well-known material systems, native $SiO_2$ on Si and $Al_2O_3$ substrates, and a Si-based AFM tip. A Hamaker constant expectation range has been determined based on literature values for both materials and compared with the Hamaker constant estimated by our methodology. Both experimentally determined Hamaker constants show a sufficient agreement with the available literature values in the right order of magnitude. The main purpose of our methodology is the fast, simple, and reliable approach to determine the Hamaker constant within an interacting system between any arbitrary-shaped particle and rough surface. The only requirement is the knowledge of one Hamaker constant of both interacting materials. In fact, the material of the AFM probe is practically always known, thus the proposed methodology can be used to determine the Hamaker constant of an unknown contamination particle. Then, if the Hamaker constant and surface topography of the particle and the underlying surface are both known, our methodology can be used to quantify the vdW force. Finally, this can potentially lead to optimized cleaning techniques, reduced downtimes, and enhanced production efficiency of EUV lithography in the semiconductor industry.

**Supporting Information**

Supporting Information is available from the Wiley Online Library or from the author.

**Acknowledgements**

The authors thank Andrey Nikipelov, Christian Cloin, Andrei Yakunin and Ksenia Makarenko of ASML for fruitful discussions regarding the experiments and Selwyn Cats (ASML) for support in coding of the vdW force model. Moreover, the authors thank Anke Aretz from Gemeinschaftslabor für Elektronenmikroskopie (GFE) of RWTH Aachen University for performing first AFM measurements and the hint towards BTR. The authors acknowledge financial support through the European Commission under the Horizon 2020 framework programme in the projects 2D Experimental Pilot Line (952792) and ZeroAMP (871749).

Received: ((will be filled in by the editorial staff))
Revised: ((will be filled in by the editorial staff))
Published online: ((will be filled in by the editorial staff))

# Supporting Information

**AFM-based Hamaker Constant Determination with Blind Tip Reconstruction**

*Benny Ku, Ferdinandus van de Wetering, Jens Bolten, Bart Stel, Mark A. van de Kerkhof, Max C. Lemme*

Hamaker Theory

In the Hamaker theory, the vdW force $F_{vdW}$ between two parallel, infinitely thick, quadratic planes with length L and separation distance d is defined by[1]

$$F_{vdW} = -\frac{A_H}{6\pi}\frac{L^2}{d^3}. \tag{1}$$

where A_H defines the Hamaker constant.

The system can be clustered into rectangular pillar elements. Then, the total $F_{vdW}$ of all cluster elements can be calculated by integration, hence introducing numerical mathematics into Equation (1). This means that the quadratic plane can be segmented into (N×N) elements with cluster element size ΔL. Then, the separation distance function d → d(x,y) becomes space dependent in the Cartesian coordinate system. The numerical vdW model can be derived to an integral

$$F_{vdW} = -\frac{A_H}{6\pi}\iint \frac{1}{d^3(x,y)} dx\, dy \approx -\frac{A_H}{6\pi}\sum_{i=1}^{N}\sum_{j=1}^{N}\frac{\Delta L^2}{d^3(x_i, y_j)}. \tag{2}$$

Through the introduction of *d(x,y)*, Equation (2) has the capability to take arbitrary body shape and surface roughness features into consideration. Thus, it can be used to compute rough particle-plane systems. The definition of *d(x,y)* is then the individual distance of any cell between particle and plane

$$d(x, y) = Z_A(x, y) - Z_B(x, y), \tag{3}$$

where $Z_A$, $Z_B$ denote the surface function of particle and plane, respectively. Additionally, *d(x,y)* requires a base line plane to define *z* = 0, which is assumed to be at $Z_B$ with min $d(x,y)$. Both bodies are then shifted along the z-axis until the minimum separation distance $d_{min} = d(x_{min},$ $y_{min}$) is reached. Therefore, the particle and plane cannot have zero distance contact since there will always be a non-zero gap due to repulsive vdW forces. For all simulations in this work, the contact distance is set to $d = d_{min} = 0.3$ nm, as it is assumed to be the averaged particle-particle distance at contact. This contact distance has been selected to be 0.3 nm, as it is representative for the vdW radii of most materials.[2] Furthermore, it make our results comparable to work of other research groups.[3–5]

Example with an Ideal Sphere-Rough Surface

For testing the vdW force model, the simplest case would be to study ideal spherical particles, as they only have one degree of freedom, the sphere radius $R$. As the simulation model is based on infinite thick cluster elements, only the lower hemisphere needs to be defined. In practical terms, most particles are sufficiently large that their body-shape above 20 nm height can be neglected. Therefore, the spherical particle is defined as lower hemispheres by

$$Z_A = -\sqrt{R^2 - X^2 - Y^2}. \tag{4}$$

Van der Waals forces have been calculated for an ideal spherical particle and artificially generated surfaces with varying ($h_{rms}$, $H$). Figure S1a and S1b show examples of such surfaces, generated with different $H$ values, but also shows that larger $H$ gives smoother surfaces. Figure S1c shows then that larger $H$ has a trend for larger $F_{vdW}$.

The results of all averaged forces $\langle F_{vdW} \rangle$ and their respective uncertainty $\sigma(\langle F_{vdW} \rangle)$ are illustrated in Figure S1c and S1d. As shown in Figure S1c, the $h_{rms}$ reduces the vdW force immensely. This is reasonable in a static model, since roughness creates voids between asperity features of a surface, which increases the volume integral in Equation (2). The results show that there are two converging plateaus which sets a lower and upper limit for $F_{vdW}$. The lower plateau exists due to several asperity contacts at finite $D$ in the simulation model. The upper plateau is the smooth sphere-plane case. This is analytically defined by $F_{vdW} = -(A_H R/6D^2)$, which agrees with the simulation result at $F_{vdW} = 11.11$ nN.

Figure S1d shows that the plateau regions as a small uncertainty with $\sigma(\langle F_{vdW} \rangle) \to 0$ nN, while the roughness cases in between both plateaus forms a ridge. It signalizes that the force histogram spread in this region is very large, which indicates that there are locally smooth, as well as rough surfaces involved in the projected area selection.

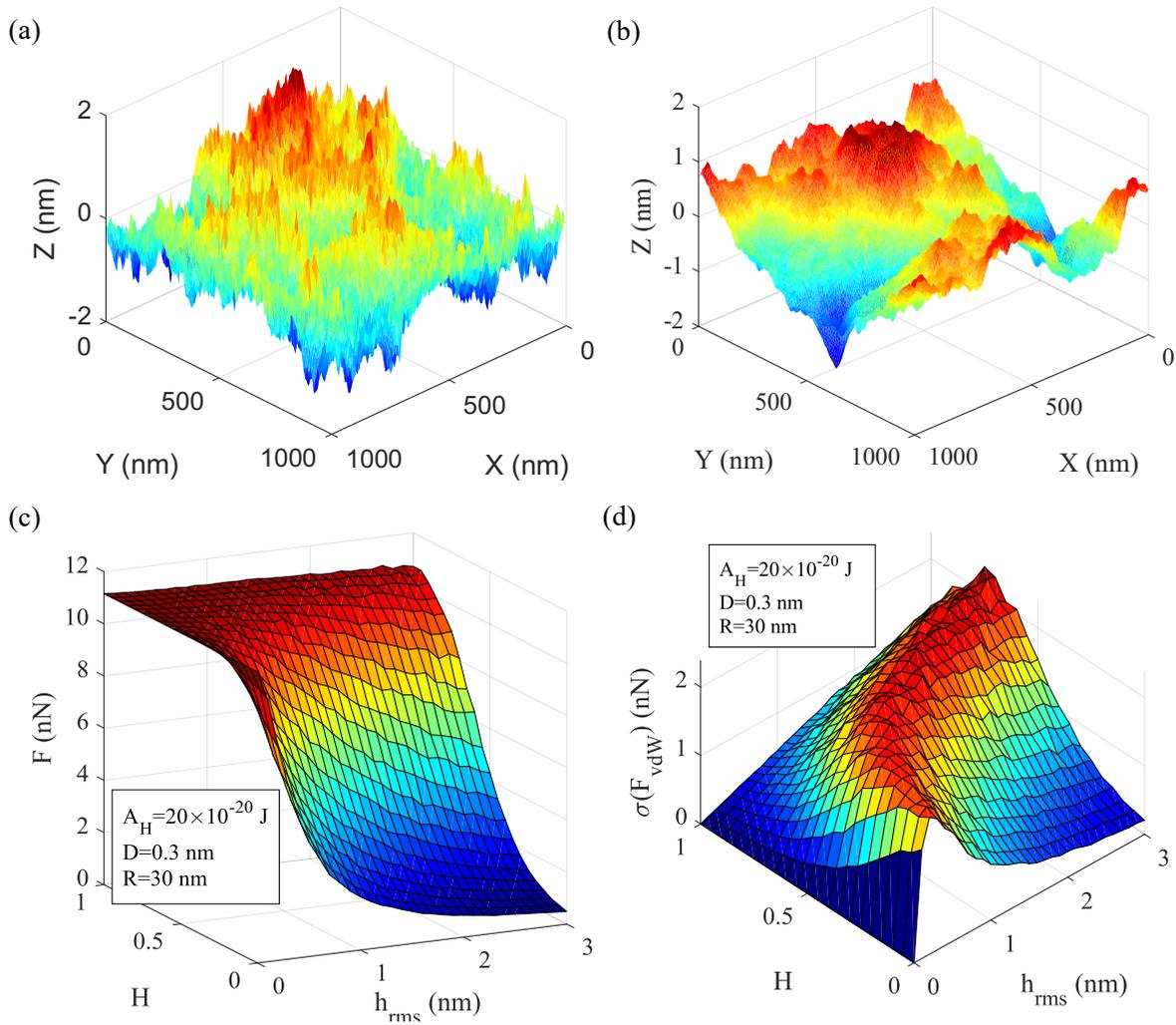

**Figure** S1. General roughness influence study of a $R = 50$ nm spherical particle at $D = 0.3$ nm within the tuning range $h_{rms} = (0-3)$ nm and $H = (0-1)$: Artificial surface example with (a) $H = 0.25$, and (b) $H = 0.75$, (c) Averaged vdW force $\langle F_{vdW} \rangle$ and its (d) standard deviation $\sigma(\langle F_{vdW} \rangle)$.

Residual Capillary Forces in Vacuum

When the cantilever approaches the sample surface, it will be snapped by vdW interactions at a distance of several nm. Once the vdW force dominates the cantilever force, the cantilever jumps into contact, which is known as snap-in force. The inverse procedure is cantilever retraction, in which the pull-off force has to be overcome to release the tip-sample contact. Even in vacuum at $10^{-4}$ mbar, capillary forces have been observed, which may be due to residual nanolayers of water from the atmosphere. This can be seen in Figure S2 by a gap between snap-in and pull-off forces. While Figure S2b shows a force difference between snap-in and pull-off in vacuum measurement is 11.58%, a force difference in the air measurement in Figure S2a amounts 47.70%. However, if there are no capillary forces, snap-in and pull-off forces should ideally match. This can sometimes be observed in the measurements, as Figure S2c and S2d show. In the end, it is difficult to make a reliable estimation about the amount of capillary forces as they could be partly included in the snap-in forces, but as well in the pull-off forces.

Usually, hot plates could be used to bake out adhered water layers, but it could also heat up the environment, what could have thermally amplified the cantilever oscillation. Therefore, a hot plate was also not used during the AFM measurements.

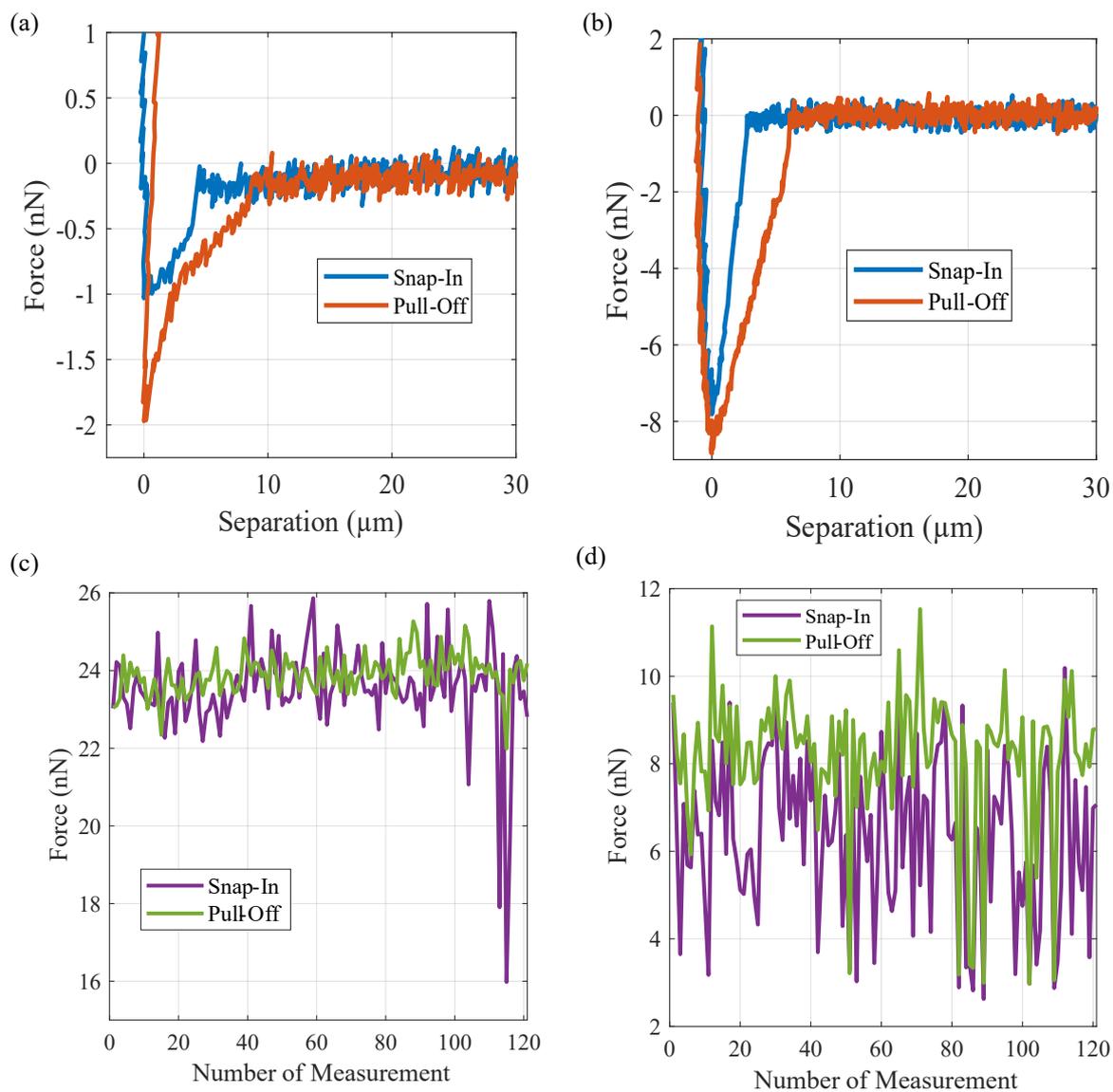

Figure S2. Characteristic force-distance curves obtained by measuring (a) in ambient conditions and (b) in vacuum. Snap-in (violet) and pull-off (green) forces of the N-th measurement from the (c) $Al_2O_3$ and (d) $SiO_2$ on Si substrates.

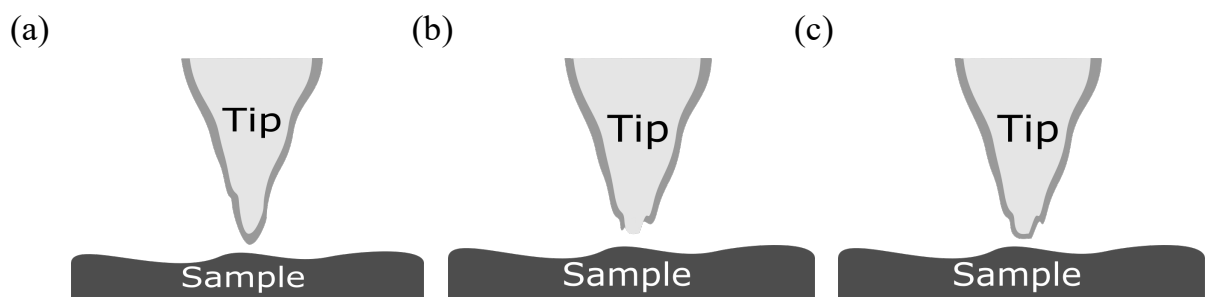

Figure S3. Tip deformation impact after contact measurement as example on a Si-based AFM tip with native oxide layer: (a) Virgin tip before contact measurement, (b) deformed tip during in-situ measurement, (c) reoxidation after air venting.

Force Calculation with an idealized sphere instead of BTR Tip

A reliable force prediction requires an accurate remodeling of the real system. This is why it is important that we suggest BTR to reconstruct the arbitrary geometrical shape of the AFM tip. Usually, the tip is approximated by an idealized geometry, such as a sphere, thus we want to show in the following example, what would happen if a sphere with R = 35 nm is used for the force calculation. We apply this example study on the experimental results of the Si tip-$Al_2O_3$ sample interaction.

Figure S4a shows the cross-section comparison between the sphere (yellow), the BTR tip (violet), and a random spot (green) on the $Al_2O_3$ sample at contact distance 0.3 nm. We selected 35 nm as the sphere radius to have a good representation with the cross-section of the BTR tip. It can also be seen that the random spot of the surface of $Al_2O_3$ sample (green) is not exactly flat. At first sight, the BTR tip and sphere may look similar in geometrical appearance, thus one might expect a similar fitting Hamaker constant for both bodies. However, the simulated force histogram (blue) in only overlays with the experimental force histogram (red), if the fitting Hamaker constant is $38 \times 10^{-20}$ J (Figure S4b). This would be not possible, as the largest Hamaker constant of the materials involved in this interaction is Si with $19 \times 10^{-20}$ J. The expectation range for the Hamaker constant is between 10 and $19 \times 10^{-20}$ J. Moreover, such high Hamaker constant would only be reasonable for metal-metal interactions.

If the surface roughness is ignored, one can also apply Equation (1) to calculate the fitting Hamaker constant for an idealized smooth sphere-plane system. If the experimental mean value 23.5 nN is assumed, the possible Hamaker constant would have been required to be $36 \times 10^{-20}$ J for a fit, which is again contradicting the literature values for the materials used in this experiment.

Therefore, one can conclude from this example that the accurate modelling of the arbitrary tip shape and the consideration of surface roughness are both very important for this case. Only the accurate geometric modelling of the real system can lead to a reasonable result of $15 \times 10^{-20}$ J, for the force interaction between an $Al_2O_3$ substrate and Si-based AFM tip as shown in the main paper of this work.

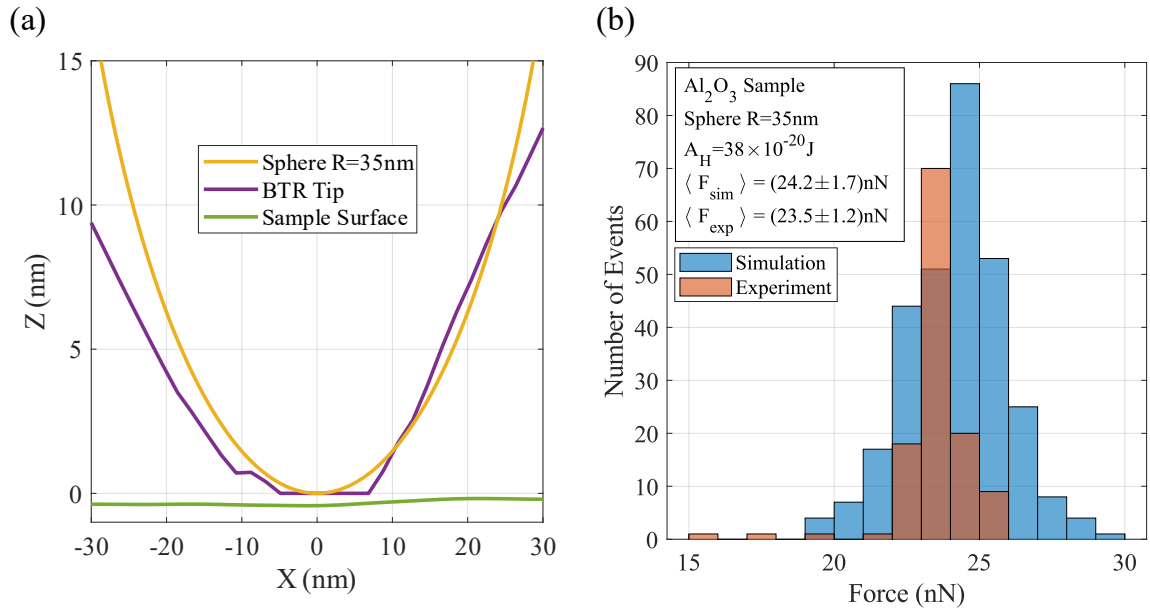

Figure S4: (a) Geometrical cross-section comparison between the sphere with radius 35 nm (yellow), BTR tip (violet) and the random spot on the $Al_2O_3$ sample surface (green). (b) Force histogram of the same sphere (blue) on the $Al_2O_3$ surface (300 different spots at a contact distance of 0.3 nm) in comparison with the measurement (red).